\documentclass{amsart}
\usepackage{hyperref}
\usepackage{amssymb}
\usepackage{amsmath}
\usepackage{pagesInBib}
\usepackage[T1]{fontenc}

\begin{document}
\newcommand{\ov}{\overline}
 \newcommand{\un}{\underline}
\renewcommand{\proof}{\bf {Proof:} \rm}
\newcommand{\BR}{{\mathbb R}}
\newcommand{\BC}{{\mathbb C}}
\newcommand{\BN}{{\mathbb N}}
\newcommand{\BZ}{{\mathbb Z}}

\renewcommand{\a}{{\bf a}}
\renewcommand{\b}{{\bf b}}
\renewcommand{\c}{{\bf c}}
\renewcommand{\d}{{\bf d}}
\newcommand{\z}{{\bf z}}

\newcommand{\p}{\underline{p}}
\newcommand{\q}{\mathfrak{q}}
\newcommand{\f}{{\bf f}}
\newcommand{\edge}{\mathfrak{e}}
\newcommand{\g}{{\bf g}}
\newcommand{\G}{{\bf G}}

\newcommand{\e}{{\bf e}}
\newcommand{\vv}{{\bf v}}
\newcommand{\bb}{{\bf b}}
\newcommand{\C}{{\bf c}}

\newcommand{\ob}{{\bf b}}
\newcommand{\cW}{{W}}
\newcommand{\cL}{{L}}
\newcommand{\cC}{{C}}
\newcommand{\cM}{{M}}

\newcommand{\cl}{C \kern -0.1em \ell}

\renewcommand{\qed}{$\blacksquare$}
\newtheorem{theorem}{Theorem}[section]
\newtheorem{remark}{Remark}[section]
\newtheorem{lemma}{Lemma}[section]
\newtheorem{proposition}{Proposition}[section]
\newtheorem{corollary}{Corollary}[section]
\newtheorem{definition}{Definition}[section]
\newtheorem{example}{Example}[section]
\newtheorem{problem}{Problem}[section]

\title[A conformal group approach to the Dirac-K\"ahler system]
{A conformal group approach to the Dirac-K\"ahler
system on the lattice}

\author{{N.~Faustino}}
\address{CMCC, Universidade Federal do ABC, 09210--580, Santo André, SP,
Brazil}
\email{\href{mailto:nelson.faustino@ufabc.edu.br}{nelson.faustino@ufabc.edu.br}}
\thanks{\href{http://orcid.org/0000-0002-9117-2021}{N.~Faustino} was formerly supported by fellowship
\href{http://www.bv.fapesp.br/14523}{13/07590-8} of FAPESP (S.P.,
Brazil)}

\date{\today}

\begin{abstract}
Starting
from the representation of the $(n-1)+n-$dimensional Lorentz
pseudo-sphere on the projective space $\mathbb{P}\mathbb{R}^{n,n}$, we propose a method to derive a class of solutions underlying to a Dirac-K\"ahler type equation on the lattice.
We make use of the Cayley transform $\varphi({\bf w})=\dfrac{1+{\bf
		w}}{1-{\bf w}}$ to show that the resulting group representation arise
from the same mathematical framework as the conformal group
representation in terms of the {\it general linear group}
$GL\left(2,\Gamma(n-1,n-1)\cup\{ 0\}\right)$. That allows us to describe such class of solutions as a commutative $n-$ary
product, involving the quasi-monomials $\varphi\left({\bf
	z}_j\right)^{-\frac{x_j}{h}}$ ($x_j \in h\mathbb{Z}$) with
membership in the paravector space $\mathbb{R}\oplus
\mathbb{R}\e_j{\bf e}_{n+j}$.
\end{abstract}



\maketitle

\section{Introduction}

Discrete function-theoretical methods has become in an emerging topic in
Clifford analysis, mainly due to the pioneering works of Faustino \&
K\"ahler (2007) \cite{FaustinoKaehler07}, Faustino et al. (2007)
\cite{FaustinoKaehlerSommen07}, De Ridder et al. (2010)
\cite{RSKS10} and the PhD dissertations of Faustino (2009)
\cite{Faustino09} and De Ridder (2013) \cite{deRidder13}. This new
research field is called {\it discrete Clifford analysis} and
corresponds to a discrete counterpart of function theory towards
the multivector representation of the null solutions for the
discretized Dirac-K\"ahler equation, in the {\it massless limit}
$m\rightarrow 0$:
\begin{eqnarray}
	\label{DiracKaehlerEq}(d-\delta)\psi=m\psi.
\end{eqnarray}

In equation (\ref{DiracKaehlerEq}) $d$ stands for the exterior
derivative, $\delta=\star^{-1} d \star$ for the co-differential form,
and $m$ for the mass term (cf.~\cite{KanamoriKawamoto04,Sushch14}).
Hereby, the symbol $\star$ represents the so-called De Rham operator.

Several approaches for finding systems of solutions associated to
discretized versions of (\ref{DiracKaehlerEq}) through combinatorics
(cf.~\cite{MalonekFalcao08,MalonekTomaz08}), Lie-algebraic
representations (cf.~\cite{RSKS10,FR11,Faustino (2013)}), and a
combination of both
(cf.~\cite{RSS12,BaaskeBRS14,FaustinoMonomiality14}) have been
worked out successfully as a unifying point of view for the theory
of orthogonal polynomials and special functions of discrete
hypercomplex variables. However, there has not been shown yet in the
context of Clifford algebras that a system of solutions for the
Dirac-K\"ahler equation on the lattice may be built up from tools of group
representation theory, although there seems to be appropriate to consider the
Poincar\'e group, the (inhomogeneous) Lorentz group and its cousins as faithful models for the study of discretized versions of the Dirac-K\"ahler equation (cf.~\cite{Lorente97,LorenteKrammer99}).

The main purpose of this paper is to construct a class of null solutions associated to a multivector discretization of the
Dirac-K\"ahler equation (\ref{DiracKaehlerEq}) on the lattice, as a
continuation of the work developed in \cite{FaustinoKGordonDirac15}. We confirm that it
can be derived in a natural way from the compactification of
$\BR^{n-1,n-1}$ at infinity by means of the Cayley
transform $\varphi({\bf w})=\dfrac{1+{\bf w}}{1-{\bf w}}$. Such
interplay indicates possibly a remarkably beautiful amalgamation
between Dirac-K\"ahler fermions on the lattice and Einstein's theory
on the Anti-de Sitter universe, yet still to be investigated in
depth such as that proposed in \cite{AFN14,Cat15}.

The paper is organized as follows:
\begin{itemize}
	\item In Section \ref{MainResultSection} we reformulate the
	construction considered in \cite{FaustinoKGordonDirac15} for an alternative discretization of the
	Dirac-K\"ahler equation (\ref{DiracKaehlerEq}). We also formulate
	the main result of this paper, Proposition \ref{MainResult}.
	\item In Section
	\ref{ConformalGroupSection} we provide the necessary background
	about Clifford algebras and the conformal group. Some references for
	this preliminary section are habilitation thesis of J.~Cnops
	(1994) \cite{Cnops94} and the research paper of V.V.~Kisil
	\cite{Kisil05} (2005); see also the books of H.B.~Lawson \&
	M.L.~Michelsohn (1989) \cite{LawsonMichels89} and W.A.~Rodrigues \&
	E.C.~de Oliveira (2007) \cite{RodriguesOliveira07} for further
	details.
	\item In Section \ref{CayleyMapSection} we make use of the mapping
	property $\mathfrak{spin}^+(n,n)\rightarrow\mbox{Spin}^+(n,n)$ to properly study the Cayley
	transform ${\bf w}\mapsto \varphi({\bf w})$. In particular, for each $j=1,2,\ldots,n$ we obtain a stereographic-like projection mapping property between the $2-$vector subspaces $\BR \e_{j}\e_{n+j}$ of $\mathfrak{spin}^+(n,n)$ and the (paravector) subspaces $H_j^{n-1,n}$ of the Lorentz pseudo-sphere $H^{n-1,n}$.
	\item In Section \ref{ProofSection} we prove
	Proposition \ref{MainResult} in detail.
	\item In Section \ref{ConclusionSection} we outlook the main
	contribution of the paper and discuss further directions of
	research.
\end{itemize}

\section{Problem Setup and Main Result}\label{MainResultSection}

The approach to be discussed throughout this paper is formulated in terms of the language of discrete multivector calculus carrying the lattice $h\BZ^n$ with meshwidth $h>0$,
and from a class of finite difference counterparts of the Dirac-K\"ahler equation (\ref{DiracKaehlerEq}), constructed from the finite difference operator
\begin{eqnarray*}
	\label{DiracEqh}
	D_h=\sum_{j=1}^n\left(\e_j\frac{\partial_h^{-j}+\partial_h^{+j}}{2}
	+\e_{n+j}\frac{\partial_h^{-j}-\partial_h^{+j}}{2}\right).
\end{eqnarray*}

The notations are the following: $\partial_h^{\pm j}$ are the
forward/backward finite difference operators
\begin{eqnarray}
	\label{DiffPmj}
	\partial_{h}^{+j}\f(x)=\dfrac{\f(x+h\e_j)-\f(x)}{h}, & \& &
	\partial_{h}^{-j}\f(x)=\dfrac{\f(x)-\f(x-h\e_j)}{h},
\end{eqnarray}
and $\e_1,\e_2,\ldots,\e_n,\e_{n+1},\ldots,\e_{2n}$ the generators of
the Clifford algebra $\cl_{n,n}$. This algebra is generated by the identity $1$, the
vectors $\e_j$, $\e_{n+k}$ $(1\leq j,k\leq n)$, and the set of anti-commuting relations
\begin{eqnarray}
	\label{CliffordBasis}
	\begin{array}{lll}
		\e_j \e_k+ \e_k \e_j=-2\delta_{jk}, & 1\leq j,k\leq n \\
		\e_{j} \e_{n+k}+ \e_{n+k} \e_{j}=0, & 1\leq j,k\leq n\\
		\e_{n+j} \e_{n+k}+ \e_{n+k} \e_{n+j}=2\delta_{jk}, & 1\leq j,k\leq
		n.
	\end{array}
\end{eqnarray}

The Clifford algebra $\cl_{n,n}$ is a linear associative algebra of dimension
$2^{2n}$, that contains the field of real numbers $\BR$ and the
Minkowski space $\BR^{n,n}$ as proper subspaces.
Here one notice that $\BR^{n,n}$ is equipped by the quadratic form
\begin{eqnarray*}
	\displaystyle Q(u,v)=\sum_{j=1}^{n}
	\left(u_j^2-v_j^2\right), &  u=(u_1,u_2,\ldots,u_{n})\in \BR^{n}&\& ~~v=(v_{1},v_{2},\ldots,v_{n})\in \BR^{n}.
\end{eqnarray*}

Next, we consider the following $\cl_{n,n}-$valued operator acting on $h\BZ^n$: $$\chi_h(x)=\prod_{j=1}^n(-1)^{\frac{x_j}{h}}\e_{n+j}\e_j.$$

This operator may be
rewritten as $\displaystyle
\chi_h(x)=(-1)^{\sum_{j=1}^n\frac{x_j}{h}} \gamma$, where
$\displaystyle \gamma=\prod_{j=1}^n\e_{n+j}\e_j$ stands for the
pseudoscalar of $\cl_{n,n}$.

The $1-$vector
representations $\displaystyle x=\sum_{j=1}^n x_j\e_j$ and $x\pm h
\e_j$ of $\BR^{n}$ will be used throughout this paper to describe
the lattice point $(x_1,x_2,\ldots,x_n)\in h\BZ^{n}$ and the
forward/backward shifts $(x_1,x_2,\ldots,x_j\pm h,\ldots, x_n)$ over
$h\BZ^n$, respectively. Also, the notation $\dfrac{\a}{\b}:=\a\b^{-1}$ will be adopted to get the
analogy with the fractional-linear transformations on the complex
plane $\BC\cong \cl_{0,1}$. Due to the non-commutativity of
$\cl_{n,n}$, one has $\dfrac{\c \a}{\c\b}\neq \dfrac{\a\c}{\b\c}$ so that only the equality $\dfrac{\a\c}{\b\c}=\dfrac{\a}{\b}$ holds
for every $\a,\b,\c$ with membership in $\cl_{n,n}$.

In \cite[Section 3]{FaustinoKGordonDirac15} it was shown that
the incorporation of the {\it local unitary action} $\chi_h(x)$ on the lattice $h\BZ^n$ allows us to
determine the null solutions of the Dirac-field operator $D_h-m\chi_h(x)$ on $h\BZ^n$ as a direct sum involving the chiral and achiral spaces of discrete multivector functions, similar to the decomposition of $\cl_{n,n}$ in terms of its even and odd parts.
Nevertheless, the primitive idempotents $\frac{1}{2}\left(1\pm \chi_h(x)\right)$ only allows us to study discretizations of (\ref{DiracKaehlerEq}) on the {\it Minkowski spacetime}, but not in a general spacetime such as the {\it conformal spacetime}.

The {\it conformal spacetime} model on the lattice was roughly discussed by Lorente \& Kramer in \cite{LorenteKrammer99} when they treat Lorentz transformations modulo rotations by means of the Cayley transform. They also have shown that the Lorentz invariance only fulfils when the lattice parameter $h$ tends to zero. However, it was not possible to fill the {\it lattice fermion doubling} on the Lorentz space when the
energy-momentum relation on the $n-$cube\footnote{The so-called $n-$dimensional {\it Brillouin zone}.} $\left[ -\frac{\pi}{h},\frac{\pi}{h}\right]^n$, as considered in \cite[Section 4]{FaustinoKGordonDirac15}:
$$\sum_{j=1}^n \frac{4}{h^2}\sin^2\left(\frac{h\xi_j}{2}
\right)=m^2$$
was replaced by its counterpart on the Lorentz space (cf.~\cite[Section 5]{LorenteKrammer99})
$$\sum_{j=1}^n \frac{4}{h^2}\tan^2\left({h\xi_j}
\right)=m^2.$$

A possible way to rid this gap was proposed by Kaplan (1992) in \cite{Kaplan92}, on which the mass $m$ was replaced by a monotonic step function. With such approach it was able to renormalize and localize the chiral anomalies in the {\it massless limit}, since the fermionic mass converges asymptotically to $\pm~m$, as the lattice parameter goes to zero. That gives in turn a valuable insight to study discrete function-theoretical methods from a mathematical physics perspective, such as the theory of finite difference potentials described on the paper \cite{CKK15}, and on the references given there.

On this paper we propose a scheme to compute the solutions underlying to a different discretization of the Dirac-K\"ahler equation. We consider for a fixed frame
$(\omega_1,\omega_2,\ldots,\omega_n)$ of the $(n-1)-$sphere
$S^{n-1}$, the following discretization
\begin{eqnarray}
	\label{DiracKaehlerEqh}
	D_h-m\omega=\sum_{j=1}^n\left(\e_j\frac{\partial_h^{-j}+\partial_h^{+j}}{2}
	+\e_{n+j}\frac{\partial_h^{-j}-\partial_h^{+j}}{2}-m\omega_j\e_{n+j}\right),
\end{eqnarray}
carrying the mass term $m$.
Hereby
$\displaystyle \omega=\sum_{j=1}^{n}\omega_j \e_{n+j}$
denotes the $1-$vector representation of $(\omega_1,\omega_2,\ldots,\omega_n)$.

From the graded anti-commuting relations (\ref{CliffordBasis}) it is straightforward to verify that
$D_h-m\omega$ satisfies the factorization property $$\left(D_h-m\omega\right)^2=\left(2hm-1\right)\Delta_h+m^2,$$
where
$
\displaystyle \Delta_h=\sum_{j=1}^n\partial_h^{-j}\partial_h^{+j}
$ denotes the {\it star-Laplacian} (cf.~\cite[p.~455]{FaustinoKaehlerSommen07}).

The factorized operator $\displaystyle \left(2hm-1\right)\Delta_h+m^2$ corresponds to a discretized version of the Klein-Gordon operator on the lattice, that differs from the one considered on \cite{FaustinoKGordonDirac15}. It has non-trivial solutions if and only if the energy-momentum condition
$$\sum_{j=1}^n \frac{4}{h^2}\sin^2\left(\frac{h\xi_j}{2}
\right)=\frac{m^2}{1-2hm}$$
is fulfilled on $\left[ -\frac{\pi}{h},\frac{\pi}{h}\right]^n$.

This paper is centered around the construction of a class of null solutions for (\ref{DiracKaehlerEqh}), through the ansatz
\begin{eqnarray}
	\label{VacuumDhm} \Psi_h(x,{\bf z})=\prod_{j=1}^n\left(\dfrac{1+{\bf
			z}_j}{1-{\bf z}_j} \right)^{-\frac{x_j}{h}},
\end{eqnarray}
whereby each term $\displaystyle {\bf z}_j \in \BR \e_j\e_{n+j}$
corresponds to a $\mathfrak{spin}^+(n,n)-$algebra representation of
the Clifford algebra $\cl_{n,n}$. That corresponds to the following proposition:

\begin{proposition}\label{MainResult}
	Let $\displaystyle \omega=\sum_{j=1}^{n}\omega_j \e_{n+j}$ be a $1-$vector representation for a point on the
	$n-$sphere $S^{n-1}$, and $\displaystyle {\bf z}=\sum_{j=1}^{n} {\bf z}_j$ a
	$\mathfrak{spin}^+(n,n)-$representation of $\cl_{n,n}$.
	Assuming that the function $\Psi_h(x,{\bf z})$ determined from the
	ansatz (\ref{VacuumDhm}) satisfies the equation $D_h\Psi_h(x,{\bf
		z})=m\omega~\Psi_h(x,{\bf z})$, we thus have the following:
	\begin{enumerate}
		\item For $m\neq 0$ the set of points ${\bf z}_j\in \BR \e_j\e_{n+j}$ $(j=1,2,\ldots,n)$ is uniquely
		determined by
		\begin{eqnarray*} \displaystyle {\bf
				z}_j=\e_j\e_{n+j}~\dfrac{\left(hm\omega_j\right)^2-2hm\omega_j}{\left(hm\omega_j\right)^2} & (j=1,2,\ldots,n).
		\end{eqnarray*}\label{zjStatement}
		\item $\Psi_h(x,{\bf z})$ equals to
		\begin{eqnarray*}\label{ClosedFormulaPsizj}
			\displaystyle \left\{\begin{array}{lll} \displaystyle \prod_{j=1}^n
				\displaystyle
				\left(\dfrac{\left(hm\omega_j\right)^2+\e_j\e_{n+j}\left(\left(hm\omega_j\right)^2-2hm\omega_j\right)}
				{\left(hm\omega_j\right)^2-\e_j\e_{n+j}\left(\left(hm\omega_j\right)^2-2hm\omega_j\right)}\right)^{-\frac{x_j}{h}}& , \mbox{for}~m \neq 0  \\
				\ \\ \displaystyle \prod_{j=1}^n
				(-1)^{\frac{x_j}{h}}
				& , \mbox{in the limit}~m\rightarrow 0
			\end{array}\right..
		\end{eqnarray*}
		Moreover, $\displaystyle \lim_{m\rightarrow 0} \Psi_h(x,{\bf z})\gamma=\chi_h(x)$, where $\gamma$ denotes the pseudoscalar of $\cl_{n,n}$, defined as above.
		\label{ClosedFormula}
		\item $\g(x)=\chi_h(x)\Psi_h(-x,{\bf z})$ is a {\it null solution} of
		$D_h-m\omega$, where $\chi_h(x)$ denotes the $\cl_{n,n}-$valued operator acting on $h\BZ^n$, defined as above.\label{VacuumStatement}
	\end{enumerate}
\end{proposition}

\section{Conformal group representation of the Lorentz pseudo-sphere}\label{ConformalGroupSection}

On this section one will introduce the conformal group representation of the $(n-1)+n-$dimensional Lorentz
pseudo-sphere 
$$
H^{n-1,n}=\left\{ (u,v)\in \BR^{n,n}~:~Q(u,v)=1\right\}
$$
of $\BR^{n,n}$ from a multivector calculus perspective.

Let us first collect some basic facts about about the Clifford algebra $\cl_{n,n}$ introduced on the previous section.
Starting from the basis graded anti-commuting relations (\ref{CliffordBasis}), one can generate the basis
elements of $\cl_{n,n}$. They consists on elements of the form
$\e_{J}=\e_{j_1}\e_{j_2}\ldots \e_{j_r}$, associated to a subset
$J=\{j_1,j_2,\ldots,j_r\}$ of $\{ 1,2,\ldots,n,n+1,\ldots,2n\}$
with cardinality $|J|=r$ so that $1\leq j_1<j_2<\ldots<j_r\leq 2n$.
For $J=\varnothing$ (empty set) one will use the convention $\e_\varnothing=1$ to denote the identity element of $\cl_{n,n}$.

Thus, any element ${\bf a}$ of $\cl_{n,n}$ may be written as
\begin{eqnarray*}
	\displaystyle \a=\sum_{r=0}^{2n}[\a]_r, & \mbox{with}&
	\displaystyle [\a]_r=\sum_{|J|=r} a_J ~\e_J.
\end{eqnarray*}

Through the projection operator $[\cdot]_r:\cl_{n,n}\rightarrow \Lambda^r (\BR^{n,n})$, the algebra $\cl_{n,n}$ can thus be associated to the following multivector
decomposition of the exterior algebra $\Lambda^*(\BR^{n,n})$ (cf.~\cite[Chapter 2]{RodriguesOliveira07}) :
$$\displaystyle \Lambda^* (\BR^{n,n})
=\bigoplus_{r=0}^{2n} \Lambda^r(\BR^{n,n}),$$
leading in particular
to the one-to-one identifications
$a\in \BR \longleftrightarrow a
\e_{\varnothing}\in \Lambda^0 (\BR^{n,n})$ and $(u,v)\in
\BR^{n,n}\longleftrightarrow {\bf u}+{\bf v}\in\Lambda^1
(\BR^{n,n})$, with
\begin{eqnarray}
	\label{Rqpvectors} \displaystyle {\bf u}=\sum_{k=1}^{n} u_k
	\e_{n+k}& \mbox{and}& \displaystyle {\bf v}=\sum_{j=1}^n v_j \e_j.
\end{eqnarray}

There is an automorphism $\a \mapsto \a'$ (main involution) and two
anti-automorphisms, $\a \mapsto \a^*$ (reversion) and $\a \mapsto
\a^\dag$ (conjugation) respectively, that leave the structure of
$\cl_{n,n}$ invariant. They are defined recursively by the rules
\begin{eqnarray}
	\label{involution}
	\begin{array}{lll}
		(\a \b)'=\a'\b' \\ (a_J \e_J)'
		=a_J~\e_{j_1}'\e_{j_2}'\ldots \e_{j_r}'~~~(1\leq j_1<j_2<\ldots<j_r\leq 2n) \\
		\e_j'=-\e_j ~~~\mbox{and}~~~ \e_{n+j}'=\e_{n+j}~~~(1\leq j\leq n)
	\end{array}
\end{eqnarray}
\begin{eqnarray}
	\label{reversion}
	\begin{array}{lll}
		(\a \b)^*=\b^*\a^* \\ (a_J \e_J)^*
		=a_J~\e_{j_r}^*\ldots \e_{j_2}^*\e_{j_1}^*~~~(1\leq j_1<j_2<\ldots<j_r\leq 2n) \\
		\e_j^*=\e_j~~~\mbox{and}~~~\e_{n+j}^*=\e_{n+j}~~~(1\leq j\leq n)
	\end{array}
\end{eqnarray}
\begin{eqnarray}
	\label{conjugation}
	\begin{array}{lll}
		(\a \b)^\dag=\b^\dag\a^\dag \\ (a_J \e_J)^\dag =a_J~\e_{j_r}^\dag
		\ldots \e_{j_2}^\dag\e_{j_1}^\dag~~~(1\leq j_1<j_2<\ldots<j_r\leq 2n) \\
		\e_j^\dag=-\e_j~~~\mbox{and}~~~\e_{n+j}^\dag=\e_{n+j}~~~(1\leq j\leq
		n).
	\end{array}
\end{eqnarray}

In particular, for a given element ${\bf u}+{\bf v}$ of
$\Lambda^1(\BR^{n,n})$, $\left({\bf u}+{\bf v}\right)'=\left({\bf
	u}+{\bf v}\right)^\dag={\bf u}-{\bf v}$ and $\left({\bf u}+{\bf
	v}\right)^*={\bf u}+{\bf v}$. Notice also that for each ${\bf
	u}+{\bf v}\in\Lambda^1(\BR^{n,n})$, $\left({\bf u}+{\bf
	v}\right)^2={\bf u}^2+{\bf v}^2$ equals to the quadratic form
$Q(u,v)$. In case where $Q(u,v)$ is non-degenerate at $(u,v)\in
\BR^{2n}$, it readily follows that
$$\left({\bf u}+{\bf v}\right)^{-1}=\dfrac{{\bf u}+{\bf v}}{{\bf
		u}^2+{\bf v}^2},$$ the so-called {\it Kelvin inverse} of ${\bf u}+{\bf v}$.

From the relations established above it
is clear that for each $n-p\geq 1$ and $n-q\geq 1$, the product of invertible $1-$vectors lying on the subspaces $\BR^{n-p,n-q}$ of $\BR^{n,n}$ remains invertible. They form the so-called {\it Clifford group}
$\Gamma(n-p,n-q)$ (cf.~\cite[Subsection 3.3.3]{RodriguesOliveira07}).

The set of all products of vectors in the Minkowski space
$\BR^{n-p,n-q}$ will be denoted by $T(n-p,n-q)$ whereas the elements ${\bf
	a}\in T(n-p,n-q)$ satisfying $\a\a^\dag=1$ form the connected subgroup
$\mbox{Pin}^+(n-p,n-q)$ of
$$ \mbox{Pin}(n-p,n-q)=\left\{ \a \in T(n-p,n-q)~:~\a\a^\dag=\pm 1\right\}.$$

For the special choice $p=q=1$ one can make use of the periodicity theorem (cf. \cite[pp. 74]{Cnops94}) to build up the correspondence between the {\it general linear group} $GL(2,\Gamma(n-1,n-1)\cup \{0\})$ of $2\times
2$ matrices, and the Clifford group
$\Gamma(n,n)$. Namely, from the one-to-one correspondence provided by the mapping
\begin{eqnarray}
	\label{PeriodicityTheorem}
	\left(\begin{array}{ccc}
		\a & \b \\
		\c & \d
	\end{array}\right) \mapsto  \dfrac{1}{2}\left[(\a+\d')+(\a-\d')\e_{2n}\e_n+(\b+\c')\e_{2n}+(-\b+\c')\right]
\end{eqnarray}
one can describe every element of $\Gamma(n,n)$ through the $2\times
2$ matrices $M=\left(\begin{array}{ccc}
\a & \b \\
\c & \d
\end{array}\right)$, whose entries $\a,\b,\c$ \& $\d$ satisfy the following conditions:
\begin{enumerate}
	\item $\a,\b,\c,\d \in \Gamma(n-1,n-1) \cup \{0\}$.
	\item $\b \d^*,\a \c^*,\a^*\b,\c^*\d \in \BR^{n-1,n-1}$.
	\item $\a \d^*-\b^*\c$ (the pseudo-determinant of $M$)
	belongs to $\BR \setminus\{0\}$.
\end{enumerate}

That corresponds to a natural extension of Waterman's approach
(cf.~\cite[Theorem 5 \& Theorem 6]{Waterman93}) to the Minkowski
space $\BR^{n-1,n-1}$. In the shed of Filmore-Springer's approach
\cite{FillmoreSpringer90}, one can define to each $M\in \Gamma(n,n)$ the representation of
$\BR^{n,n}$ the {\it conformal group}
$\mathcal{M}(n-1,n-1)$ of $\overline{\BR^{n-1,n-1}}:=\BR^{n-1,n-1}\cup \{
\infty\}$\footnote{The notation $\overline{\BR^{n-1,n-1}}:=\BR^{n-1,n-1}\cup
	\{ \infty\}$ means the compactification of $\BR^{n-1,n-1}$ by the point
	at infinity.}, the so-called M\"obius transformations on $\BR^{n-1,n-1}$
(cf.~\cite[Subsection 5.1]{Cnops94}):
$$
\mu_M : {\bf z} \mapsto \dfrac{\a \z + \b}{\c \z +\d}.
$$

It is well known that for every $M \in GL\left(2,\Gamma(n-1,n-1)\cup
\{0 \}\right)$, the {\it pseudo-orthogonal group}\footnote{$O(n,n)$
	is the group of isometries that preserves $Q(\cdot,\cdot)$ so that
	$SO(n,n)$ acts as an isometry on $H^{n-1,n}$.} $O(n,n)$ may be
conformally embedded on the projective space $\mathbb{P}\BR^{n,n}$
through the mapping $S\mapsto MS(M')^{-1}$, whereby $\displaystyle
{S}=\left(\begin{array}{ccl}
0 & 1  \\
-1  & 0
\end{array}\right)$ denotes a $2 \times 2$ representation of
$H^{n-1,n}$ for the inversion mapping ${\bf
	w}\mapsto -{\bf w}^{-1}$ on $\cl_{n,n}$
(cf.~\cite[p.~93]{Waterman93}).

By employing the involution map (\ref{involution}) and the isomorphism (\ref{PeriodicityTheorem}),
there holds $M'=\left(\begin{array}{ccl}
{\bf a}' & -{\bf b}'  \\
-{\bf c}'  & {\bf d}'
\end{array}\right)$ (cf.~\cite[p.~75]{Cnops94}). This leads to 
\begin{eqnarray*}
	MS=\left(\begin{array}{ccc}
		-\b & \a \\
		-\d & \c
	\end{array}\right) & \mbox{and}& SM'=\left(\begin{array}{ccl}
		-{\bf c}' & {\bf d}'  \\
		-{\bf a}'  & {\bf b}'
	\end{array}\right).
\end{eqnarray*}

Therefore, the equation $MS(M')^{-1}=S$ is fulfilled whenever $M$ is of the form
\begin{eqnarray}
	\label{SpinGroup} M=\left(\begin{array}{ccl}
		{\bf a} & {\bf b}  \\
		{\bf b}'  & {\bf a}'
	\end{array}\right),
\end{eqnarray}
\mbox{with} ${\bf a},{\bf b}\in T(n-1,n-1)~,{\bf a}{\bf
	b}^*\in\BR^{n-1,n-1}$ and $|{\bf a}|^2-|{\bf b}|^2=1$.

Mimicking \cite[Proposition 2.2.]{Kisil05}, the family of M\"obius
transformations that preserve the Lorentz pseudo-sphere $H^{n-1,n}$ are of the form
$$
\mu_M' : {\bf z} \mapsto \dfrac{\a \z + \b}{\b' \z +\a'}.
$$
They will be denoted throughout by $\mathcal{M}^+(n-1,n-1)$.

\section{Compactification through the Cayley
	map}\label{CayleyMapSection}

It is worth mentioning that, unlike to the sphere\footnote{$S^{n-1}$
	is an affine subspace of $H^{n-1,n}$.} $S^{n-1}$ the Lorentz pseudo-sphere
$H^{n-1,n}$ does not possess a group structure, and hence one cannot
identify, as in case of $S^{n-1}$, the manifold $H^{n-1,n}$ with the left coset
$K\setminus \mathcal{M}^+(n-1,n-1)$, involving the maximal compact
subgroup (cf.~\cite[pp.~742-743]{Kisil05})
$$
K=\left\{ \left(\begin{array}{ccl}
\dfrac{{\bf u}}{|{\bf u}|} & {\bf 0}  \\
{\bf 0}  & \dfrac{{\bf u}'}{|{\bf u}|}\end{array}\right)~:~{\bf
	u}\in \Gamma(n-1,n-1)\right\}.
$$

However $\mbox{Pin}^+(n,n)$ gives a double covering of
$\mathcal{M}^+(n-1,n-1)$ (cf.~\cite[Subsection 5.3]{Cnops94}), so
that $H^{n-1,n}$ may be embedded in a group manifold of
$\mathcal{M}^+(n-1,n-1)$. There is an alternative and more useful
description for the conformal group $\mathcal{M}^+(n-1,n-1)$ that
starts with the identification of the points of the Minkowski space
$\BR^{n-1,n-1}$ as elements with membership in a certain Lie
algebra, on which the passage from the Lie algebra to the Lie group
by means of the Cayley map
\begin{eqnarray}
	\label{CayleyMap} \varphi({\bf w})=\dfrac{1+{\bf w}}{1-{\bf w}}
\end{eqnarray}
yields the compactification $\overline{\BR^{n-1,n-1}}$.

Here one
notice that the associated Cayley transform (\ref{CayleyMap}) is
encoded by the $\mbox{Spin}^+(n,n)-$representation $$\displaystyle
\dfrac{1}{\sqrt{2}}(I+S)=\left(\begin{array}{ccl}
\frac{1}{\sqrt{2}} & \frac{1}{\sqrt{2}}  \\
-\frac{1}{\sqrt{2}}  & \frac{1}{\sqrt{2}}\end{array}\right).$$

We easily check that
\begin{eqnarray}
	\label{CayleyMapPM}
	\begin{array}{lll}
		\varphi({\bf w})+\varphi({\bf w})^{-1}&=& 2\dfrac{1+{\bf
				w}^2}{1-{\bf
				w}^2} \\
		\varphi({\bf w})-\varphi({\bf w})^{-1}&=& \dfrac{4{\bf w}}{1-{\bf
				w}^2}\\
		{\bf w}(\varphi({\bf w})+\varphi({\bf w})^{-1}+2)&=&\varphi({\bf w})-\varphi({\bf
			w})^{-1},
	\end{array}
\end{eqnarray}
hold for every ${\bf w}^2\neq 1$, that is
$\frac{1}{2}\left(\varphi({\bf w})\pm\varphi({\bf w})^{-1}\right)$
may be interpreted as coordinates of a stereographic-like
projection.

In his habilitation thesis \cite{Cnops94} J. Cnops (1994) obtained,
in particular, a generalization of the above description in terms of
the Clifford group $\Gamma(n,n)$, following the insights of Filmore
and Springer \cite{FillmoreSpringer90} (see \cite[Chapter
5]{Cnops94}). In the shed of Cnop's characterization provided by
\cite[Corollary 5.2.9]{Cnops94}, the description of $\Gamma(n,n)$
may be represented in terms of $2\times 2$-matrix representations of
the group $GL(2,\Gamma(n-1,n-1)\cup\{ 0\})$, as despicted on Section \ref{ConformalGroupSection}. Since
$\mbox{Pin}^+(n,n)$ is a subgroup of $\Gamma(n,n)$, it suffices (as
it will be seen next) to describe $\overline{\BR^{n-1,n-1}}$ in
terms of the spin group $\mbox{Spin}^+(n,n)=\mbox{Pin}^+(n,n)\cap
\cl_{n,n}^0$, where
$$
\cl_{n,n}^0 =\{ \a \in \cl_{n,n}~:~\a'=\a\}
$$
corresponds to the even subalgebra of $\cl_{n,n}$.

Next, we examine the action of the Cayley map (\ref{CayleyMap}) on
the (paravector) subspaces $H_j^{n-1,n}$ of $H^{n-1,n}$, defined as
$$H_j^{n-1,n}=\left\{~v_j+u_j\e_j\e_{n+j}\in
\BR \oplus\BR\e_j\e_{n+j}~:~~v_j>0,~v_j^2-u_j^2=1\right\}.$$

Here we recall that $\mathfrak{spin}^+(n,n)=\mathfrak{spin}(n,n)$
coincides with the space of $2-$vectors~(cf.~\cite[Proposition
6.1]{LawsonMichels89}) $$\displaystyle
\Lambda^2(\BR^{n,n})=\mbox{span}\left\{~\e_j\e_k~:~1\leq j<k\leq
2n\right\}$$
so that $\displaystyle
\frac{1}{\sqrt{2}}(I+S):~\mathfrak{spin}^+(n,n) \rightarrow
\mbox{Spin}^+(n,n)$ yields the compactification
$\overline{\BR^{n-1,n-1}}$ in terms of the action ${\bf w}\in
\Lambda^2(\BR^{n,n}) \mapsto \varphi({\bf w})$. The restriction of
the Cayley map (\ref{CayleyMap}) to the subspaces $\BR\e_j\e_{n+j}$
($j=1,2,\ldots,n$) of $\Lambda^2(\BR^{n,n})$ will be of special
interest in the proof of Proposition \ref{MainResult}, mainly
because for the (paravector) subspaces $H_j^{n-1,n}$ of $H^{n-1,n}$ the map $\varphi~:~\BR\e_j\e_{n+j}\rightarrow H_j^{n-1,n}$ is
evidently onto. This can be easily deduced from the
stereographic-like parametrization
\begin{eqnarray}
	\label{CayleyStereographic}\left\{
	\begin{array}{lllll}
		v_j&=&\dfrac{1+{\bf z}_j^2}{1-{\bf z}_j^2} \\
		& & &,~{\bf z}_j\in \BR\e_j\e_{n+j}
		\\
		u_j\e_j\e_{n+j}&=&\dfrac{2{\bf z}_j}{1-{\bf z}_j^2}
	\end{array}\right..
\end{eqnarray}

\section{Proof of Proposition \ref{MainResult}}\label{ProofSection}

It is clear from (\ref{CayleyMap}) and (\ref{CayleyStereographic})
that for every $\mathfrak{spin}^+(n,n)$-representation of the form
$\displaystyle {\bf z}=\sum_{j=1}^n {\bf z}_j$ $({\bf
	z}_j\in\BR \e_j\e_{n+j})$, and for every $x\in h\BZ^n$, the function $\Psi_h(x,{\bf z})$ may be rewritten as
\begin{eqnarray}
	\label{n-aryProduct} \Psi_h(x,{\bf z})=\prod_{j=1}^n \varphi({\bf
		z}_j)^{-\frac{x_j}{h}}=\prod_{j=1}^n
	\left(v_j+u_j\e_j\e_{n+j}\right)^{-\frac{x_j}{h}}.
\end{eqnarray}

This function corresponds to a commutative $n-$ary product. From the
anti-commuting relations (\ref{CliffordBasis}), it is clear that $2-$vectors
$\e_{j}\e_{n+j}$ and $\e_{k}\e_{n+k}$ commute
($k=1,2,\ldots,n$; $j \neq k$ fixed). Thus, for every $x_j,x_k\in
h\BZ$, one readily has
\begin{eqnarray*}
	\left(v_k+u_k\e_k\e_{n+k}\right)^{-\frac{x_k}{h}}
	\left(v_j+u_j\e_j\e_{n+j}\right)^{-\frac{x_j}{h}}=\left(v_j+u_j\e_j\e_{n+j}\right)^{-\frac{x_j}{h}}
	\left(v_j+u_j\e_j\e_{n+j}\right)^{-\frac{x_j}{h}},
\end{eqnarray*}
that is $\varphi({\bf z}_k)^{-\frac{x_k}{h}}\varphi({\bf
	z}_j)^{-\frac{x_j}{h}}=\varphi({\bf z}_j)^{-\frac{x_j}{h}}\varphi({\bf
	z}_k)^{-\frac{x_k}{h}}$.

From now on, we assume that
$\Psi_h(x,{\bf z})$ is a solution of the discrete Dirac equation
$D_h\g(x)=m\omega~\g(x)$. From (\ref{DiffPmj}) and (\ref{DiracKaehlerEqh}),
this is equivalent to
\begin{eqnarray*}
	\label{DiracKaehlerEqPsih}
	\begin{array}{lll}
		\displaystyle \sum_{j=1}^n \left(\e_j~\frac{\Psi_h(x+h\e_j,{\bf
				z})-\Psi_h(x-h\e_j,{\bf z})}{2h} -
		\e_{n+j}\frac{\Psi_h(x+h\e_j,{\bf z})+\Psi_h(x-h\e_j,{\bf
				z})}{2h}\right)={\bf m}_h \Psi_h(x,{\bf z}),
	\end{array}
\end{eqnarray*}
with $\displaystyle {\bf m}_h(\omega)=\sum_{j=1}^n\left(m \omega_j-
\frac{1}{h}\right)\e_{n+j}.$

Straightforward computations based on (\ref{n-aryProduct}) gives
rise to the set of finite difference equations
\begin{eqnarray*}\displaystyle
	\left\{\begin{array}{lll}\displaystyle \e_j\frac{\Psi_h(x+h\e_j,{\bf
				z})-\Psi_h(x-h\e_j,{\bf z})}{2h} &=&\displaystyle
		-\frac{1}{h}\e_j\dfrac{2{\bf z}_j}{1-{\bf z}_j^2}\Psi_h(x,{\bf z})
		\\ \ \\
		\displaystyle\e_{n+j}\frac{\Psi_h(x+h\e_j,{\bf
				z})+\Psi_h(x-h\e_j,{\bf z})}{2h} &=& \displaystyle
		-\frac{1}{h}\e_{n+j}~ \frac{1+{\bf z}_j^2}{1-{\bf
				z}_j^2}\Psi_h(x,{\bf z})
	\end{array}\right. & (j=1,2,\ldots,n).
\end{eqnarray*}

Then, from the set of properties
\begin{eqnarray*}
	\left(\dfrac{1+{\bf z}_j^2}{1-{\bf
			z}_j^2}\right)^2-\left(\dfrac{2{\bf z}_j}{1-{\bf z}_j^2}\right)^2=1
	& \mbox{for}& j=1,2,\ldots,n,
\end{eqnarray*}
the finding of the solution for the equation $D_h\Psi_h(x,{\bf
	z})=m\omega\Psi_h(x,{\bf z})$ reduces to the problem of finding the set of
points ${\bf z}_j\in \BR \e_j\e_{n+j}$ such that the system of
equations
\begin{eqnarray*}\label{VacuumDhmPhizj}
	\left\{\begin{array}{lll} \displaystyle -\e_j\dfrac{2{\bf
				z}_j}{1-{\bf z}_j^2}&=&\displaystyle
		\frac{1}{2}\e_{n+j}\left[\left(hm\omega_j- 1\right)-
		\left(hm\omega_j- 1\right)^{-1}\right]  \\
		\ \\ \displaystyle \e_{n+j}\dfrac{1+{\bf z}_j^2}{1-{\bf
				z}_j^2}&=& \displaystyle \frac{1}{2}\e_{n+j}\left[\left(hm\omega_j- 1\right)+ \left(hm\omega_j- 1\right)^{-1}\right]
	\end{array}\right. & (j=1,2,\ldots,n)
\end{eqnarray*}
is fulfilled for each $\displaystyle \omega=\sum_{j=1}^{n}\omega_j \e_{n+j}$ with membership in $S^{n-1}$.

From the above system of equations, the terms $\varphi({\bf z}_j)$
of the right-hand of (\ref{n-aryProduct}) are determined from the set
of equations
\begin{eqnarray}
	\label{varphizj}
	\begin{array}{lll}
		\varphi({\bf z}_j)&=& \displaystyle \dfrac{1+{\bf z}_j^2}{1-{\bf
				z}_j^2}+\dfrac{2{\bf z}_j}{1-{\bf
				z}_j^2}\nonumber \\
		\nonumber \\ &=& \displaystyle \dfrac{1}{2}\left[\left(hm\omega_j- 1\right)+ \left(hm\omega_j- 1\right)^{-1}\right]+\dfrac{1}{2}\e_j\e_{n+j}\left[\left(hm\omega_j- 1\right)- \left(hm\omega_j- 1\right)^{-1}\right].
	\end{array}
\end{eqnarray}

\subsection*{Proof of Statement (\ref{zjStatement}) of Proposition
	\ref{MainResult}}

From the equation $\displaystyle {\bf z}_j\left(\dfrac{1+{\bf
		z}_j^2}{1-{\bf z}_j^2}+1\right)=\dfrac{2{\bf z}_j}{1-{\bf z}_j^2}$
provided by (\ref{CayleyMapPM}), one readilly has
\begin{eqnarray*}
	{\bf z}_j&=&\e_j\e_{n+j}~\dfrac{\dfrac{1}{2}\left[\left(hm\omega_j- 1\right)- \left(hm\omega_j- 1\right)^{-1}\right]}{\dfrac{1}{2}\left[\left(hm\omega_j- 1\right)+ \left(hm\omega_j- 1\right)^{-1}\right]+1}.
\end{eqnarray*}

A short computation
gives rise to
\begin{eqnarray*}
	{\bf z}_j=\e_j\e_{n+j}~\dfrac{\left(hm\omega_j- 1\right)^2-
		1}{\left(hm\omega_j- 1\right)^2+2\left(hm\omega_j-1\right)+1} &
	(j=1,2,\ldots,n),
\end{eqnarray*}

or equivalently, to the simplified formula
\begin{eqnarray*} \displaystyle {\bf
		z}_j=\e_j\e_{n+j}~\dfrac{\left(hm\omega_j\right)^2-2hm\omega_j}{\left(hm\omega_j\right)^2} & (j=1,2,\ldots,n).
\end{eqnarray*}

This corresponds to the set of points for which
$D_h\Psi(x,{\bf z})=m \omega~\Psi_h(x,{\bf z})$ is fulfilled, as desired.

\subsection*{Proof of Statement (\ref{ClosedFormula}) of Proposition \ref{MainResult}}

In case where $m\neq 0$, the ${\bf z}_j'$s determined as above satisfy
${\bf z}_j \neq -\e_j\e_{n+j}$ so that $\displaystyle \varphi({\bf
	z}_j)=\frac{1+{\bf z}_j^2}{1-{\bf z}_j^2}+\frac{2{\bf z}_j}{1-{\bf
		z}_j^2}$ is well defined (see eq.~(\ref{CayleyMapPM})).

From a short computation, involving the identity $\displaystyle
\frac{1-{\bf a}{\bf b}^{-1}}{1+{\bf a}{\bf b}^{-1}}=\frac{{\bf
		b}-{\bf a}}{{\bf b}+{\bf a}}$ there holds
\begin{eqnarray*} \displaystyle \varphi({\bf
		z}_j)=\dfrac{\left(hm\omega_j\right)^2+\e_j\e_{n+j}\left(\left(hm\omega_j\right)^2-2hm\omega_j\right)}
	{\left(hm\omega_j\right)^2-\e_j\e_{n+j}\left(\left(hm\omega_j\right)^2-2hm\omega_j\right)} & (j=1,2,\ldots,n).
\end{eqnarray*}

In case where $m\rightarrow 0$, it readily follows from equation (\ref{varphizj}) that
$$\varphi({\bf z}_j)=-1.$$

Finally, by inserting the $\varphi({\bf z}_j)'$s on the
right-hand side of (\ref{n-aryProduct}), we finish the {\bf proof of
	Statement (\ref{ClosedFormula})}.

\subsection*{Proof of Statement (\ref{VacuumStatement}) of Proposition \ref{MainResult}}

First, we recall that the operator $$\displaystyle
\chi_h(x)=\prod_{j=1}^n(-1)^{\frac{x_j}{h}}\e_{n+j}\e_j$$ satisfy
the unitary property, $\chi_h(x)^2=1$, the set of recursive
equations $\chi_h(x\pm h\e_j)=-\chi_h(x)$ ($j=1,2,\ldots,n$) and the
set of graded anti-commuting relations
\begin{eqnarray*}
	\chi_h(x)\e_j+\e_j\chi_h(x)=
	\chi_h(x)\e_{n+j}+\e_{n+j}\chi_h(x)=0 & \mbox{for} &
	j=1,2,\ldots,n.
\end{eqnarray*}

Then, from (\ref{DiffPmj}) and (\ref{DiracKaehlerEqh}) the function $\f(x)$ is a null solution for
$D_h-m\omega$ if and only if the function $\f(x)$ satisfies the equation
\begin{eqnarray*} \sum_{j=1}^n\left(-
	\e_j~\chi_h(x)\frac{\f(x+h\e_j)-\f(x-h\e_j)}{2h}
	-\e_{n+j}\chi_h(x)\frac{2\f(x)-\f(x+h\e_j)-\f(x-h\e_j)}{2h}\right)=
	-m\omega ~\chi_h(x)\f(x).
\end{eqnarray*}

Hence, for $\g(x)=\chi_h(x)\f(x)$ the above equation is equivalent
to
\begin{eqnarray*} \sum_{j=1}^n\left(
	\e_j~\frac{\g(x+h\e_j)-\g(x-h\e_j)}{2h}
	-\e_{n+j}\frac{\g(x+h\e_j)+\g(x-h\e_j)}{2h}\right)=-{\bf
		m}_h(\omega) \g(x),
\end{eqnarray*}
with $\displaystyle {\bf m}_h(\omega)=\sum_{j=1}^n\left(m\omega_j-
\frac{1}{h}\right)\e_{n+j}.$

Putting $\f(x)=\Psi_h(-x,{\bf z})$, with $\displaystyle {\bf
	z}=\sum_{j=1}^n{\bf z}_j$ (${\bf z}_j\in\BR \e_j\e_{n+j}$),
there holds from a straightforward computation that the solution of
the above equation is determined from the system of equations
(\ref{VacuumDhmPhizj}). Therefore, one can easily infer that $\g(x)=\chi_h(x)\Psi_h(-x,{\bf z})$ is a solution of the equation
$D_h\g(x)=m\omega\g(x)$, as desired.

\section{Discussion}\label{ConclusionSection}

A scenario of modifying the lattice discretization of the Dirac-K\"ahler equation (\ref{DiracKaehlerEq}) was proposed throughout this paper. Two ingredients towards such modification were crucial. One of them consists on the treatment of the mass term $m$
as a $1-$vector term of the Clifford algebra of signature $(n,0)$, say $m\omega$ with $\omega\in S^{n-1}$. The other was the construction of a one-to-one mapping
between the subspaces $\BR\e_j\e_{n+j}$ of  $\mathfrak{spin}^+(n,n)$ and the
(paravector) subspaces $H_j^{n-1,n}$ of the Lorentz pseudo-sphere $H^{n-1,n}$, by means of the stereographic-like projection (\ref{CayleyStereographic}). 
That was the main key on the proof of Proposition
\ref{MainResult}.

We believe that
the same technique may also be applied to non-compact symmetric
spaces that admit a horospherical or Iwasawa decomposition such as the {\it Einstein static universe}, carrying the universal cover of $\overline{\BR^{n-1,n-1}}$ ~(cf.~\cite{GibbonsSteif95}). 

What we have tried to show throughout this paper is that the conformal group is encoded on the discretization of Dirac operators considered in a series of papers, and by several authors in the context of {\it discrete Clifford analysis}. This feature is not tangible if we only consider projection operators to decouple the discrete Dirac-K\"ahler equation on its components (cf. \cite{Sushch14}).

Although this paper offers a different perspective to study discretizations of the Dirac-K\"ahler equation (\ref{DiracKaehlerEq}), the problem of {\it lattice fermion doubling} regarding the chiral operator $\chi_h(x)$ in $h\BZ^n$ was not duly clarified. 
Here we would like to stress that $\displaystyle \chi_h(x)=\lim_{m\rightarrow 0}\Psi_h(x,{\bf z})\gamma$ is not a null solution of the discrete Dirac operator $D_h$. 

One possible way to overcome the aforementioned problem may consists on the replacement of the mass term $m$ by the ratio $\dfrac{m}{\Lambda_h}$, where $\Lambda_h$ stands a {\it cutoff term}.
It should be recalled that in the case of the terms\footnote{The resulting components of the right-hand side of (\ref{varphizj}) obtained from the change of variable $m \rightarrow \dfrac{m}{\Lambda_h}$.} $\dfrac{\omega_j}{\Lambda_h}$ 
of
\begin{eqnarray*}
	\exp\left(\e_j\e_{n+j}\log\left(h\frac{m}{\Lambda_h}~\omega_j- 1\right)\right)&=&\dfrac{1}{2}\left[\left(h\frac{m}{\Lambda_h}~\omega_j- 1\right)+ \left(h\frac{m}{\Lambda_h}\omega_j- 1\right)^{-1}\right]\\
	&+&\dfrac{1}{2}\e_j\e_{n+j}\left[\left(h\frac{m}{\Lambda_h}\omega_j- 1\right)- \left(h\frac{m}{\Lambda_h}\omega_j- 1\right)^{-1}\right]
\end{eqnarray*}
satisfy the asymptotic condition $\frac{\omega_j}{\Lambda_h}\sim \frac{2}{hm}$, as $m\rightarrow 0$, one has $$\displaystyle
\exp\left(\e_j\e_{n+j}\log\left(h\frac{m}{\Lambda_h}~\omega_j- 1\right)\right)\sim 1$$ 
in the {\it massless limit}. This is what Kaplan (1992) already did on the paper \cite{Kaplan92} when he flipped the chirality gap by introducing a fermionic mass, parametrized in terms of hyperbolic coordinates.

There are other important questions that also deserve to be investigated as a
whole and whether these results can be exploited in the shed of the Anti-De Sitter spacetime. 
In such model, the lattice term $\dfrac{m}{\Lambda_h}$ that appears on the eigenvalue-type equation $\displaystyle D_h \f(x)=\dfrac{m}{\Lambda_h}\omega \f(x)$, involving the discrete Dirac operator $D_h$, may play the role of the cosmological constant in Einstein's equation. 

We conjecture that the introduction of the {\it cutoff term} $\Lambda_h$ in our model will allow us to obtain the complete picture for the null solutions of $D_h$, near the conformal infinity. A case of special interest will be when the componentwise terms $\dfrac{m}{\Lambda_h}\omega_j$ of $D_h-\dfrac{m}{\Lambda_h}\omega$ satisfy the following asymptotic expansion, on the limit $\Lambda_h\rightarrow \pm \infty$:
\begin{eqnarray*}
	\dfrac{m}{\Lambda_h}\omega_j \sim \dfrac{1-\cosh(h\xi_j)}{h}, &\mbox{for}&
	\cosh(h\xi_j)=\dfrac{1-{\bf z}_j^2}{1+{\bf z}_j^2}.
\end{eqnarray*}

In case that such condition is fulfilled, one gets an asymptotic expansion for the symmetric part of $D_h$, i.e.  $\displaystyle D_h-\dfrac{m}{\Lambda_h}\omega \sim \sum_{j=1}^{n}\e_j \frac{\partial_{h}^{-j}+\partial_{h}^{+j}}{2}$(cf.~\cite[Section 4]{FaustinoKGordonDirac15}). Such characterization, if feasible, will provide an outstanding step on the theory of {\it discrete monogenic functions}.

\subsection*{Acknowledgments}

The author would like to thank to the anonymous referees for the carefully reading and their help in the revision of the preliminary version of this paper.


\end{document}